\documentclass[manuscript]{aastex701}
\usepackage{amsmath}

\renewcommand{\arraystretch}{1.5}
\usepackage{chngcntr}
\counterwithin{table}{section}

\begin{document}

\title{Global Multi-ion Solar Wind Model. I. Ion Temperatures}

\author[0000-0001-5260-3944]{Bart van der Holst}
\affiliation{Astronomy Department, Boston University, Boston, MA 02215, USA}
\email[show]{bartvand@bu.edu}

\author[0000-0002-9465-7470]{Judit Szente}
\affiliation{Astronomy Department, Boston University, Boston, MA 02215, USA}
\email{judithsz@bu.edu}

\author[0000-0002-9325-9884]{Enrico Landi}
\affiliation{Climate and Space Sciences and Engineering Department, University of Michigan, Ann Arbor, MI 48109, USA}
\email{elandi@umich.edu}

\begin{abstract}
Over the past several decades, observations have shown that minor ions have a higher temperature and flow faster than protons in the solar wind. Theories based on turbulence have been developed that can explain many of these observed phenomena. We present our first step in developing a global multi-ion solar wind model with turbulence by including ion temperatures but not yet including differential streaming. The extent of this model is from the lower transition region (50,000 K temperature) to the corona and inner heliosphere. It uses low-frequency, reflection-driven incompressible turbulence to address coronal heating and solar wind acceleration. The energy partitioning of the turbulence dissipation to the electrons and various ions is based on stochastic heating and linear Landau and transit-time damping. In order to test the validity of our approach we have carried out a three-dimensional simulation of the solar corona and the solar wind using an idealized dipole magnetic field configuration, calculated the Oxygen temperature across the entire domain, and compared it to measurements obtained from the UltraViolet Coronagraph Spectrometer (UVCS) on the Solar and Heliospheric Observatory (SOHO) satellite and with the Solar Wind Ion Composition Spectrometer (SWICS) on board Advanced Composition Explorer (ACE). The comparison shows that even with the simplified magnetic field configuration the multi-ion model predictions reproduce the heavy-ion preferential heating phenomena in both remote-sensing and in-situ observations.
\end{abstract}


\section{Introduction} 

The solar atmosphere plays a fundamental role in the solar-terrestrial relationship, because it is the source of the two main outputs which influence the near-Earth space, the Earth's upper atmosphere, and are responsible for large scale disturbances. In fact, the solar atmosphere emits high energy radiation (X-rays, EUV and UV) as well as particles and magnetic field in the form of the continuously streaming solar wind \citep{Cranmer:2019,Verscharen:2019}. These types of output are subject to sudden and very large increases when flares occur -- releasing large amounts of high energy radiation -- and when Coronal Mass Ejections are launched into space \citep{Forbes:2000}. The latter carry a large quantity of mass along with embedded magnetic field, both of which can interact with planetary magnetospheres and cause geomagnetic storms \citep{Tsurutani:2009,Liu:2021}: these storms can disrupt communications, pose significant hazard for human assets in space (including astronauts), and damage the power distribution infrastructure on the ground.

In recent decades, a large body of work has tried to identify the physical processes generating flares and CMEs, and to develop tools and methods to predict them \citep{Kataoka:2009,Singh:2020,Shen:2021,Riley:2025,vanderholst:2025}. However, predicting capabilities are still insufficient to provide enough lead time and geo-effectiveness estimates, and no definitive consensus has been reached concerning the physical processes that heat and accelerate solar coronal plasma into the solar wind, that generate flares, and that are responsible for causing entire large scale solar magnetic structures to erupt into the heliosphere. As part of this effort, a number of first principles 3D models of the entire solar atmosphere and heliosphere have been built, e.g.\ Alfv\'en Wave Solar atmosphere Model \citep[AWSoM,][]{vanderHolst:2014,vanderHolst:2022}, Magnetohydrodynamic Algorithm outside a Sphere \citep[MAS,][]{Mikic:2018,Lionello:2023}, European Heliospheric Forecasting Information Asset \citep[EUHFORIA,][]{Pomoell:2018}, with the aim of predicting the properties of the solar atmosphere and the solar wind during quiescence and activity, as well as studying the processes occurring in solar and heliospheric plasmas.

Solar plasmas are mainly composed of Hydrogen, with some significant amount of Helium (\citet{moses:2020} reported Helium coronal abundance, given as the ratio of number densities of $\approx$8-10\% in the open field line region between 1.3 and $3.0\,R_\odot$) and only trace amounts of all other elements -- the minor heavy ions -- the most abundant of which make less than 0.01\% of the total number of particles. Still, these less abundant elements are responsible for the bulk of the high energy emission of the solar atmosphere, which is observed remotely by space instruments, and they have been routinely detected by in-situ instruments in the last few decades. Furthermore, minor heavy ion physical properties (e.g., ionization status, kinetic energy, bulk speed, relative abundances, velocity distribution functions) provide some of the most important signatures of the physical processes responsible for solar particle and radiation output both in quiescence and during active phase \citep{abbo:2016}.

Despite their importance for diagnostic use, the calculation of the observed properties of minor ions has never been fully developed and integrated into 3D models. In particular, the inclusion of minor ions in 3D global models is limited to calculation of charge states \citep{Lionello:2019, Szente:2022}. Recently, \citet{Shi:2019} and \citet{Szente:2023} demonstrated that such departure from ionization equilibrium both in the magnetically closed corona and most importantly in the solar wind strongly affects the spectral analysis; \citet{Szente:2023} utilized such an out-of-equilibrium charge state distribution to calculate the spectrum emitted by the solar atmosphere using the SPECTRUM module in AWSoM developed by \citet{Szente:2019}, demonstrating that departures from the commonly adopted ionization equilibrium assumption are ubiquitous and significantly alter solar emission. \citet{Wraback:2025,wraback:2026} used AWSoM to model the charge state evolution of a CME from launch at the Sun to Earth, showing that depending on the particular processes CME plasmas were subject to during their early acceleration the CME charge state composition is extremely diverse on very small spatial scales and is a very powerful diagnostic for the properties that generated the measured charge state distribution.

However, determining the charge state distribution of the plasma in a 3D model is relatively straightforward because collisional ionization and recombination can be modeled using the plasma electron temperature, electron density, and speed everywhere in the model's domain calculated without including heavy ions in the model's energy and momentum equations. On the other hand, kinetic properties such as heavy ion bulk speed, as  well as thermal and non-thermal velocity distributions, require a multi-ion approach. This paper, the first of a series, has two goals: 1) describes such an implementation on the AWSoM model and 2) presents a simulation that includes all Oxygen ions using an idealized dipole magnetic configuration, and we compare the ion temperatures of protons, O$^{5+}$, O$^{6+}$, and O$^{7+}$ with in-situ measurements from the ACE/SWICS time of flight spectrometer \citep{Gloeckler:1998SSRv...86..497G} and from spectroscopic measurements of widths of the O$^{5+}$ lines observed by the SoHO/UVCS spectrometer \citep{Kohl:1995,Li:1998,Esser:1999}. Future papers in this series will 1) utilize a realistic photospheric magnetic field map to predict spectroscopic and in-situ signatures of ion temperatures from an array of instruments, to fully assess the quality of the model predictions as well as to investigate the diagnostic potential of this implementation, and then 2) they will be expanded to include differential acceleration, comparing model results to observations.

Minor ions have already been included in previous work: \citet{Isenberg:1982,Isenberg:1984} developed an expression for wave turbulence in multi-ion component plasma and demonstrated the heating and acceleration of $\alpha$-particles, protons, and electrons in a 1D model. \citet{Buergi:1986,Buergi:1992} developed a 1D model for the heating and acceleration of protons, $\alpha$-particles, and heavy ions in the solar wind \citet{Habbal:1995} included alpha particles in the determination of the solar wind flow properties using a two-fluid model with constraints based on white light images and in-situ data. \citet{Hansteen:1997} investigated the role of $\alpha$-particles in the outer solar atmosphere without turbulence. \citet{LieSvendsen:2003} generalized this to a 1D gyrotropic model for the corona and the solar wind. \citet{Endeve:2005} again generalized this 1D model to include expansion factors to model various open- and closed-field regions of the Sun. \citet{Li:2008} developed a 2D steady state model of the multi-ion component solar wind plasma with Alfv\'en wave turbulence. \citet{Ofman:2015} was the first to develop a three-dimensional multi-ion solar corona model, but without turbulence. In our paper, we develop a 3D multi-ion solar corona and solar wind model from the lower transition region to 1AU that includes incompressible turbulence.

This paper is organized as follows: Section~\ref{model} discusses the physical processes included in the multi-ion AWSoM model, while their implementation is described in Section~\ref{implementation}. Model predictions are reported in Section~\ref{results}, and their comparison with observations is discussed in Section~\ref{observations}. This work is summarized in Section~\ref{summary}.

\section{Computational Model}
\label{model}

The present multi-ion global solar wind model is a single fluid MHD model, with different densities and temperatures for each ion species. In Section \ref{sec:mhd}, we describe this MHD system. In Section \ref{sec:turbulence}, we discuss the incompressible turbulence that is used to address coronal heating and solar wind acceleration. The energy partition of the dissipated turbulence to the various ion species and electrons is detailed in Section \ref{sec:partition}.

\subsection{Governing Equations of Multi-ion Model}\label{sec:mhd}

The starting point of our multi-ion model is the MHD equations, which describe the gross-macroscopic properties of the solar wind plasma. The time evolution of the mass densities is given by
\begin{equation}
  \frac{\partial \rho_i}{\partial t} + \nabla\cdot(\rho_i{\bf u}) = \frac{\delta\rho_i}{\delta t},
\end{equation}
where $\rho_i$ is the mass density of the ion species $i$ and ${\bf u}$ is the bulk velocity, assumed to be the same for the electrons and all ion species. The source term on the right hand side is due to ionization and recombination processes:
\begin{equation}
    \frac{\delta\rho_i}{\delta t} = m_in_e\left\{n(X^{m-1})C_{m-1}\left(T_e\right) - n(X^m) \left[C_{m}\left(T_e\right) + R_{m}\left(T_e\right)\right] + n(X^{m+1}) R_{m+1}\left(T_e\right)\right\}
    \label{eq:ionreceomb}
\end{equation}
in which we have identified the ion $i$ with mass $m_i$ as element $X_m$ in charge state $m$ and number density $n(X^m)$. The electron number density can be obtained from charge neutrality
\begin{equation}
  n_e = \frac{1}{e}\sum_i q_i n_i,
\end{equation}
in which $n_i$ and $q_i$ are the ion number density and ion charge and $e$ is the elementary charge. The coefficients $C_m(T_e)$ and $R_m(T_e)$ describe the total ionization and recombination rates and are obtained from CHIANTI \citep{2024ApJ...974...71D}, similar to \citet{Szente:2022}. These coefficients depend on the electron temperature $T_e$. The conservation of momentum is given by
\begin{align}
  \frac{\partial \rho{\bf u}}{\partial t}
  &+ \nabla\cdot(\rho{\bf u}{\bf u}) + \nabla p + \nabla p_e
  - {\bf j}\times{\bf B} + \nabla\cdot{\bf P}_w \nonumber \\
  & = - \rho \left[ \frac{GM_\odot}{r^3}{\bf r}+{\bf\Omega}\times({\bf\Omega}\times{\bf r}) + 2{\bf\Omega}\times{\bf u}\right],
\end{align}
where
\begin{equation}
    \rho=\sum_i\rho_i,\qquad p=\sum_i p_i,
\end{equation}
are the total ion mass density and pressure, $p_e$ is the electron pressure, ${\bf B}$ is the magnetic field, ${\bf j} = \nabla\times{\bf B}/\mu_0$ is the current density, $\mu_0$ is the permeability of vacuum, $G$ is the gravitational constant, $M_\odot$ is the solar mass, ${\bf r}$ is the position vector relative to the center of the Sun and ${\bf\Omega}$ is the angular velocity of the Sun. We assume a uniform solar rotation with a 25.38 days period so that $\Omega = 2.865\times10^{-6}\;{\rm rad}\,{\rm s}^{-1}$. ${\bf P}_w$ is the turbulence pressure tensor consisting of the Reynolds-averaged stress tensor and turbulent pressure, which will be discussed in Section \ref{sec:turbulence}. The ion and electron pressures are determined by the equations
\begin{align}
  \frac{\partial p_i}{\partial t}
  &+ \nabla\cdot\left( p_i{\bf u}\right)
  + (\gamma-1)p_i\nabla\cdot{\bf u}
  = \frac{\delta p_{i}}{\delta t} + (\gamma-1)Q_{i},
  \label{eq:ionpressure} \\
  \frac{\partial p_e}{\partial t}
  &+ \nabla\cdot\left( p_e{\bf u}\right)
  + (\gamma-1)p_e\nabla\cdot{\bf u}=
  \frac{\delta p_e}{\delta t} + (\gamma-1)(Q_e-Q_{\rm rad}-\nabla\cdot{\bf q}_e),
\end{align}
where $\gamma = 5/3$. The Coulomb collisional source terms in the electron and ion pressure are:
\begin{equation}
    \frac{\delta p_s}{\delta t} = \sum_t n_s\mu_{st}\nu_{st}\left[
                \frac{2k_B(T_t - T_s)}{m_t}  \right].
\end{equation}
The indices $s$ and $t$ loop over all ion species $i$ and electron $e$. The temperature $T_s$ of species $s$ is determined by the equation-of-state $p_s=n_sk_BT_s$, where $k_B$ is the Boltzmann constant. Furthermore, the collision frequencies are
\begin{equation}
        \nu_{st} = \ln\Lambda\frac{\sqrt{\mu_{st}}}{m_s}\left(\frac{q_sq_t}{\varepsilon_0}\right)^2\frac{1}{3(2\pi k_B)^{3/2}} \frac{n_t}{T_{st}^{3/2}},
\end{equation}
where $\mu_{st} = m_sm_t/(m_s+m_t)$ is the reduced mass and $T_{st} = (m_tT_s+m_sT_t)/(m_s+m_t)$ is the reduced temperature, $\ln\Lambda\approx20$ is the Coulomb logarithm, $\varepsilon_0$ is the permittivity of vacuum. The electron and ion coronal heating functions are denoted by $Q_e$ and $Q_i$, respectively. Their sum is equal to the total dissipation of the turbulence described in Section \ref{sec:turbulence} with the partitioning detailed in Section \ref{sec:partition}. The optically thin radiative loss in the lower corona is
\begin{equation}
  Q_{\rm rad} = n_en_p  L(T_e),
\end{equation}
where $L(T_e)$ is the radiative cooling curve obtained from CHIANTI. The electron heat flux ${\bf q}_e$ consists of two parts. Near the Sun, we use the collisional formulation of Spitzer:
\begin{equation}
  {\bf q}_{e,S} = -\kappa_e T_e^{5/2}{\bf b}{\bf b}\cdot\nabla T_e,
\end{equation}
where ${\bf b} = {\bf B}/B$ and
$\kappa_e \approx 9.2\times 10^{-12}\;{\rm W}\,{\rm m}^{-1}\,{\rm K}^{-7/2}$,
while further away from the Sun, we use the collisionless heat flux as suggested by \cite{Hollweg:1978}:
\begin{equation}
  {\bf q}_{e,H} = \frac32 \alpha p_e {\bf u},
\end{equation}
in which we assume $\alpha=1.05$. We smoothly transition between these formulations
\begin{equation}\label{eq:heatflux}
  {\bf q}_e = f_S {\bf q}_{e,S} + (1 - f_S){\bf q}_{e,H}.
\end{equation}
Here, the fraction of Spitzer heat flux is defined as a function of $r$
\begin{equation}\label{eq:interpolation}
  f_S = \frac{1}{1 + (r/r_H)^2},
\end{equation}
where $r_H = 5R_\odot$ similar to \citet{Chandran:2011}. We did not include source/sink terms due to recombination and ionization, since the energy involved in ionizing/recombining minor ions is very small compared to other energies, except when protons and to some extent Helium are ionizing/recombining (and they are ionizing at very low temperatures), due to the low abundance of any element beyond H and He. The magnetic field is obtained from the induction equation
\begin{equation}
  \frac{\partial {\bf B}}{\partial t}
  - \nabla\times({\bf u}\times{\bf B}) = 0,
\end{equation}
with the constraint of $\nabla\cdot{\bf B}=0$ for the initial condition.

\subsection{Incompressible Turbulence}\label{sec:turbulence}

Here, we give a short description of the incompressible turbulence we use in our model. A more complete turbulence model and derivations can be found in, for example, \citet{Wang:2022} and references therein.

The starting point of incompressible turbulence is to split the magnetic field and velocity vectors as sums of regular and turbulent parts, ${\bf u} = \tilde{\bf u} + \delta {\bf u}$ and ${\bf B} = \tilde{\bf B} + \delta {\bf B}$ (below tildes are omitted) and assume that the turbulent amplitudes satisfy the incompressibility conditions:
\begin{equation}
    \nabla\cdot\delta{\bf u} =0, \quad {\bf B}\cdot\delta{\bf B}=0, \quad \nabla\cdot\delta{\bf B} =0.
\end{equation}
Instead of velocity and magnetic field fluctuations, we will use  Els\"asser amplitudes ${\bf z}_\pm = \delta{\bf u}\pm\delta {\bf B}/\sqrt{\mu_0\rho}$ and define the associated energy densities:
\begin{equation}
    w_\pm = \frac{1}{4}\rho{\bf z}^2_\mp, \qquad w_D = \frac{1}{2}\rho{\bf z}_-\cdot{\bf z}_+.
\end{equation}
Here, the energy related to the correlator ${\bf z}_-\cdot{\bf z}_+$ is essentially a difference between the fluctuating kinetic and magnetic energy: $w_D=\frac{1}{2}\rho\delta u^2 - \frac{1}{2\mu_0}\delta B^2$. The turbulence energy densities are then obtained from the following time evolution equations:
\begin{align}
    \frac{\partial w_\pm}{\partial t} &+ \nabla\cdot\left[ w_\pm (
    {\bf u}\pm{\bf V}_{\rm A})\right] + \frac{1}{2}w_\pm\nabla\cdot{\bf u}
    = - (S \pm R)\frac{w_D}{2} -\epsilon_\pm w_\pm, \label{eq:wplusminus}\\
 \frac{\partial w_D}{\partial t} &+ \nabla\cdot(w_D{\bf u})+\frac{1}{2}w_D\nabla\cdot{\bf u} = -(S-R)w_+-(S+R)w_- -\frac{1}{2}(\epsilon_++\epsilon_-)w_D, \label{eq:wdifference}
\end{align}
The Alfv\'en velocity uses the total ion mass density ${\bf V}_A={\bf B}/\sqrt{\mu_0\rho}$. For transverse turbulence, the shear flow and turbulence reflection source terms are defined via
\begin{equation}
    S=\frac{1}{2}\nabla\cdot{\bf u} - {\bf b}\cdot({\bf b}\cdot\nabla){\bf u},
    \qquad R = ({\bf b}\cdot\nabla)V_A,
\end{equation}
respectively. The last term on the right-hand side of Equation \ref{eq:wplusminus} is the dissipation rate. In our model, we use the phenomenological cascade rate of \citet{Dmitruk:2002},
\begin{equation}
    \epsilon_\pm = \frac{2}{L_\perp\sqrt{\rho}}\sqrt{w_\mp}
\end{equation}
which contains the transverse correlation length $L_\perp$ of the turbulence in the plane perpendicular to the magnetic field ${\bf B}$. We prescribe $L_\perp$ similarly to \citet{Hollweg:1986} by setting $L_\perp\sqrt{B}$ as a constant. To avoid that the reflected turbulence energy density is larger than the outward propagating turbulence energy density, we limit the Alfv\'en speed gradient reflection coefficient by $\left| R \right| \leq \frac{1}{2}\left| \epsilon_+ - \epsilon_- \right|$. The third term on the left-hand side of Equation \ref{eq:wplusminus} is the reduction in turbulence energy due to expanding flows and combined with the shear flow source term on the right-hand side are due to work done by the turbulence pressure tensor
\begin{equation}
    {\bf P}_{\rm w} = p_{{\rm w}\perp} I +
    (p_{{\rm w}\parallel}-p_{{\rm w}\perp}) {\bf b}{\bf b},
\end{equation}
where
$p_{{\rm w}\perp} = \frac{1}{2}(w + w_D)$ and $p_{{\rm w}\parallel} = \frac{1}{2}(w - w_D)$ are the perpendicular and parallel turbulent pressure, respectively, and $w=w_++w_-$ is the total turbulent energy density.

The equations for the ions, electrons, magnetic field, and turbulence can be recast in a near conservation form for the total energy density
\begin{align}
    \frac{\partial E}{\partial t}
  &+\nabla\cdot\left[ \left(E + p + p_e + \frac{B^2}{2\mu_0}\right){\bf u} +{\bf P}_{\rm w}\cdot{\bf u}
  - \frac{\mathbf{u}\cdot\mathbf B}{\mu_0}\mathbf{B}  
  + (w_+ - w_-){\bf V}_A \right] \nonumber \\
  & = -\nabla\cdot{\bf q}_e -Q_{\rm rad}- \rho{\bf u}\cdot \left[\frac{GM_\odot}{r^3}{\bf r}+{\bf\Omega}\times({\bf\Omega}\times{\bf r})\right],
\end{align}
where
\begin{equation}
    E = \frac{\rho u^2}{2} + \frac{p}{\gamma -1} + \frac{p_e}{\gamma-1} +\frac{B^2}{2\mu_0} + w,
\end{equation}
is the total energy density. Here, we assumed that
\begin{equation}
    \epsilon_+w_++\epsilon_-w_- = Q_e + \sum_iQ_i,
\end{equation}
i.e., all dissipated turbulence energy is channeled to the coronal heating of the electrons and ions.

\subsection{Energy Partitioning}\label{sec:partition}

In this Section, we present a generalization to multiple ion species of the energy partition introduced by \citet{Chandran:2011} in a 1D solar wind model and used in the 3D solar wind model of \citet{vanderHolst:2014,vanderHolst:2022}. This energy partition assumes that the non-linear interaction of counter-propagating turbulence energy densities results in a transverse energy cascade from the outer scale through the self-similar inertial range to the proton gyro-radius scale $r_p$ where the Alfv\'enic cascade transitions into kinetic Alfv\'en wave (KAW) cascade. \citet{Chandran:2011} made the approximation that KAW dissipation occurs on two distinct scales: $\sim r_p$ and $\ll r_p$. Around the proton gyroradius scale, they calculated damping rates associated with several key damping mechanisms: stochastic heating, linear Landau damping, and linear transit time damping. The remaining cascade power that did not dissipate at $r_p$ contributes only to electron heating.

We now derive the cascade and dissipation process for multiple ion species. \citet{Boldyrev:2005} showed that the cascade time of the Els\"asser variables ${\bf z}_{\pm,\lambda}$ at length scale $\lambda$ is
\begin{equation}\label{eq:cascadetime}
  \tau_{\pm,\lambda} = \frac{\lambda}{z_{\mp,\lambda} \sin(\theta_\lambda)},
\end{equation}
where $\theta_\lambda$ is the alignment angle between ${\bf z}_{-,\lambda}$ and ${\bf z}_{+,\lambda}$. We assume that the transverse energy cascade rate in the inertial range is independent of $\lambda$
\begin{equation}\label{eq:cascadepower}
  Q_{\pm,\lambda} = \frac{w_{\pm,\lambda}}{\tau_{\pm,\lambda}}
    \propto \lambda^0=1,
\end{equation}
so that by using Equations \ref{eq:cascadetime} and \ref{eq:cascadepower} the following identity is obtained:
\begin{equation}
  \frac{Q_{\pm,\lambda}}{Q_{\pm,{L_\perp}}} = 1
  = \frac{w_{\pm,\lambda}}{w_{\pm,{L_\perp}}}
  \sqrt{\frac{w_{\mp,\lambda}}{w_{\mp,{L_\perp}}}}
  \frac{L_\perp \sin(\theta_\lambda)}{\lambda \sin(\theta_{L_\perp})}.
\end{equation}
where $\theta_{L_\perp}$ is the alignment angle at length scale $\lambda=L_\perp$. Under the assumption of small angle $\theta_\lambda$ and using the angle scaling $\theta_\lambda\propto\lambda^{1/4}$ derived in \citet{Boldyrev:2005}, we obtain the following
\begin{equation}
  \frac{w_{\pm,\lambda}}{w_{\pm,{L_\perp}}}
  \sqrt{\frac{w_{\mp,\lambda}}{w_{\mp,{L_\perp}}}}\left(\frac{L_\perp}{\lambda}
  \right)^{3/4} = 1.
\end{equation}
The turbulence energy densities are then estimated as
\begin{equation}
  w_{\pm,\lambda} = w_{\pm,{L_\perp}} \sqrt{\frac{\lambda}{L_\perp}}
\end{equation}
at length scale $\lambda$ in the inertial range.

In the dissipation range, the cascaded power will decrease, since some of the cascaded energy dissipates at $k_\perp r_i \sim 1$, where $r_i = V_{\perp i }/\Omega_i$ is the ion gyro radius, and $V_{\perp i} = \sqrt{2k_BT_{\perp i}/m_i}$ is the perpendicular thermal speed and $\Omega_i = (q_i/m_i)B$ is the ion-cyclotron frequency. In the following, we assume that the ions are ordered such that the ion gyro-radius increases with index $i$. Then for ions $i<N$, where $N$ is the total number of ion species considered, we have
\begin{equation}
    Q_{\pm,r_i} = (1-\Gamma_{\pm,r_{i+1}})Q_{\pm,r_{i+1}},
\end{equation}
where $\Gamma_{\pm,r_i}$ is the fraction of cascaded power dissipated at $k_\perp r_i\sim 1$. Assuming that in between subsequent gyro-radii there is no dissipation and using Equations \ref{eq:cascadetime} and \ref{eq:cascadepower} for the cascade time and cascade power, we obtain the following:
\begin{equation}
    \frac{Q_{\pm,r_i}}{Q_{\pm,r_{i+1}}} = 1-\Gamma_{\pm,r_{i+1}} = 
    \frac{w_{\pm,r_i}\sqrt{w_{\mp,r_i}}r_{i+1}\sin(\theta_{r_i})}
    {w_{\pm,r_{i+1}}\sqrt{w_{\mp,r_{i+1}}}r_i\sin(\theta_{r_{i+1}})}.
\end{equation}
We also assume that the alignment angle $\theta_\lambda$ is small, so that $\sin(\theta_{r_i})\approx \theta_{r_i}$. By defining 
\begin{align}
    F &= \frac{w_{+,r_i}}{w_{+,r_{i+1}}}, \qquad G = \frac{w_{-,r_i}}{w_{-,r_{i+1}}}, \label{eq:FG} \\
    \sigma &= \frac{r_{i+1}\sin(\theta_{r_i})}{r_i\sin(\theta_{r_{i+1}})},    
\end{align}
and using the angle scaling derived in \citet{Boldyrev:2005}, i.e., $\theta_\lambda \propto \lambda^{1/4}$, so that
\begin{equation}
    \sigma = \left( \frac{r_{i+1}}{r_i} \right)^{3/4},
\end{equation}
we arrive at
\begin{equation}
    1-\Gamma_{+,r_{i+1}} = F\sqrt{G}\sigma, \qquad 1-\Gamma_{-,r_{i+1}} = G\sqrt{F}\sigma.
\end{equation}
From which we obtain
\begin{align}
    F &= \frac{\left(1-\Gamma_{+,r_{i+1}}\right)^{4/3}}
        {\left(1-\Gamma_{-,r_{i+1}}\right)^{2/3}}\left( \frac{r_i}{r_{i+1}}\right)^{1/2},\\
    G &= \frac{\left(1-\Gamma_{-,r_{i+1}}\right)^{4/3}}
        {\left(1-\Gamma_{+,r_{i+1}}\right)^{2/3}}\left( \frac{r_i}{r_{i+1}}\right)^{1/2},
\end{align}
which then provides $w_{\pm,r_i}$ from $w_{\pm,r_{i+1}}$ via Equation \ref{eq:FG}.

For stochastic heating of the ion $i$ at gyro-radius $r_i$, we first need to determine the velocity fluctuation. By assuming equipartition at the ion gyro-scale, we obtain the following
\begin{equation}
    \rho\delta u_{r_i}^2 = w_{r_i}, \qquad \delta B_{r_i}^2/\mu_0 = w_{r_i},
\end{equation}
where
\begin{equation}
    w_{r_i} = w_{+,r_i} + w_{-,r_i},
\end{equation}
is the total wave energy density at the ion gyro-scale. The effective damping rate at $k_\perp r_i\sim 1$ and
$\beta_{\parallel p}=2\mu_0 p_{\parallel p}/B^2 \lesssim 1$ from stochastic ion heating is \citep{Chandran:2010}
\begin{equation}
  \gamma_{\perp i} = \frac{Q_{\text{stoch}, i }}{w_{r_i}}
  =c_1 \frac{\rho_i \delta u_{r_i}^3}{r_i w_{r_i}}
  \exp\left( -\frac{c_2}{\varepsilon_i}\right)
\end{equation}
where $c_1$ and $c_2$ are dimensionless constants of order unity and $\varepsilon_i = \delta u_{r_i}/V_{\perp i}$ is a normalized velocity fluctuation. Although we currently do not include differential streaming in our model, we note however that it will have a strong impact on the damping rate; see \citet{Chandran:2013}. Together with the cascade time at the ion gyro scale $r_i$
\begin{equation}
    \tau_{\pm,r_i} = \frac{w_{\pm,r_i}}{Q_{\pm,r_i}},
\end{equation}
we can then define the fraction of cascade power dissipated at gyro-radius $r_i$
\begin{equation}
  \Gamma_{\pm,{r_i}} = \frac{\gamma_{r_i}\tau_{\pm,r_i}}
        {1+\gamma_{r_i}\tau_{\pm,r_i}}, \label{eq:dampingrate}
\end{equation}
where
\begin{equation} \label{eq:heatingfraction}
    \gamma_{r_p} = \gamma_{\parallel e} + \gamma_{\perp p}+\sum_{i=1}^N \gamma_{\parallel i}, \qquad
    \gamma_{r_i} = \gamma_{\perp i}, \qquad \mbox{for }i=2,\cdots,N,
\end{equation}
are the damping rates at the proton gyro-radius $r_p$ and the minor ion gyro-radius $r_i$, respectively. The damping rates $\gamma_{\parallel e}$ and $\gamma_{\parallel p}$ at the proton gyro scale are described in \cite{vanderHolst:2022} and use the critical-balance condition \citet{Lithwick:2007,Goldreich:1995}. In the following, we set $\gamma_{\parallel i}=0$ for $i=2,\cdots,N$. In future work, the latter can be obtained, for instance, from the PLUME model \citep{Klein:2015,Klein:2025} to solve the contribution of minor ions to the linear Landau and linear transit time damping rates of kinetic Alfv\'en waves at the proton gyro-radius.

The final heating functions are expressed with the damping rates and heating fractions, Equations \ref{eq:dampingrate} and \ref{eq:heatingfraction}, respectively:
\begin{align}
    Q_{\perp i} &= Q_{r_i}, \qquad i=2,\cdots,N,\\
    Q_{\perp p} &= \frac{\gamma_{\perp p}}{\gamma_{r_p}}Q_{r_p},
    \qquad Q_{\parallel i} = \frac{\gamma_{\parallel i}}{\gamma_{r_p}}Q_{r_p}, \qquad i=1,\cdots,N, \\
    Q_e &= \frac{\gamma_{\parallel e}}{\gamma_{r_p}}Q_{r_p} + (Q_{+,{r_p}} + Q_{-,{r_p}} - Q_{r_p}), \label{eq:eheat}
\end{align}
where we defined for $i=1,\cdots,N$:
\begin{align}
    Q_{+,{r_i}} &= \epsilon_+w_+\prod_{j=i+1}^N(1-\Gamma_{+,r_j}), \\
    Q_{-,{r_i}} &= \epsilon_-w_-\prod_{j=i+1}^N(1-\Gamma_{-,r_j}), \\
    Q_{r_i} &= \Gamma_{+,r_i}Q_{+,{r_i}} + \Gamma_{-,r_i}Q_{-,{r_i}}.
\end{align}
The last term in brackets in Equation \ref{eq:eheat} is due to the remaining cascading power that cascades further to $k_\perp\gg r_p$ and is assumed to contribute only to electron heating.

In the following, we assume that the abundance of minor ions is negligible since we will not include $\alpha$ particles in this paper. We also assume that the fraction of cascade power dissipated at $k_\perp r_i\sim 1$ is negligible. Then for all minor ions, $i=2,\cdots,N$, we obtain
\begin{equation}
    w_{\pm,r_i} = w_\pm\sqrt{\frac{r_i}{L_\perp}}.
\end{equation}
In this case, we no longer need to sort the ions according to their gyro radius in every spatial location and at every time step. The coronal heating functions for the minor ions can be approximated by
\begin{equation}
    Q_{\perp i} = \Gamma_{+,r_i}\epsilon_+w_+ + \Gamma_{-,r_i}\epsilon_-w_-,
\end{equation}
and
\begin{equation}
    Q_{+,{r_p}} = \epsilon_+w_+\left( 1 - \sum_{i=2}^N \Gamma_{+,{r_i}}\right), \qquad
    Q_{-,{r_p}} = \epsilon_-w_-\left( 1 - \sum_{i=2}^N \Gamma_{-,{r_i}}\right).
\end{equation}
The total coronal heating for each ion is $Q_i = Q_{\perp i} + Q_{\parallel i}$. Note that since we set $\gamma_{\parallel,i}=0$, we have $Q_{\parallel,i}=0$ for $i=2,\cdots,N$. We also note that very close to the Sun the oxygen gyro radii can be smaller than the proton gyro radius, but there the electron heating is the most dominant heating.

\section{Model Implementation}
\label{implementation}

In this section, we present some details of the implementation of the global multi-ion solar wind model. The starting point for the model implementation is the Alfv\'en Wave Solar atmosphere Model \citep[AWSoM,][]{vanderHolst:2014} which is part of the overarching Space Weather Modeling Framework \citep[SWMF,][]{Toth:2012}.

The inner boundary conditions are set in the lower transition region. The temperatures all have the same value of $T_e=T_i=50,000\,$K. We overestimate the proton density for this temperature by an order of magnitude setting it at the value $n_p = 2\times 10^{17}\,$m$^{-3}$. With this density overestimation, radiative cooling will be significant. By maintaining a temperature floor of $50,000\,$K the density will fall radially outward until radiative cooling is no longer able to drop the temperature below $50,000\,$K. From that radius outward, the temperature will rise above $50,000\,$K. The densities of the ions at the inner boundary are set by assuming a zero ionization and recombination source term, Equation \ref{eq:ionreceomb}, and assuming coronal abundances \citep{Feldman:1992}. The radial magnetic field component is prescribed either by a synoptic magnetic field map or by setting an analytic solution for the initial magnetic field. For the test in this paper we use a dipole field with $5.6\,$G field strength at the magnetic pole. The dipole has a tilt of $15^\circ$. For the longitudinal and latitudinal field components we use an extrapolation with zero gradient of the difference with the initial magnetic field, i.e. in the test below the variation with the dipole field. For the velocity boundary conditions, we use field aligned extrapolation, while the other velocity components are set to zero in the frame co-rotating with the Sun.

The energy density of the outbound Alfv\'en waves is prescribed via the Poynting flux $S_A$: $w=(S_A/B)_\odot\sqrt{\mu_0\rho}$, where $B$ and $\rho$ are the field strength and the total mass density at the inner boundary. For the test in this paper, we set $(S_A/B)_\odot = 10^6\,$W\,m$^{-2}$\,T$^{-1}$. The inward propagating turbulence energy density and the difference energy density, $w_D$, are set to zero at the inner boundary. Additionally, we set the correlation length to be inversely proportional to $\sqrt{B}$: $L_\perp\sqrt{B}=1.5\times10^5\,$m\,$\sqrt{\rm T}$.

The computational grid is a 3D spherical grid that extends over the ranges $R_\odot<r<24\,R_\odot$, $0\leq\theta\leq\pi$, and $0\leq\varphi<2\pi$. The grid is stretched towards the Sun, with the smallest radial grid cell size of $\Delta r\approx 2\times 10^{-4}\,R_\odot$. We use adaptive mesh refinement (AMR) with blocks of $6\times8\times8$ internal mesh cells. The setup of AMR is such that between $1.3\,R_\odot$ and $1.7\,R_\odot$ the angular resolution is $1^\circ.4$, between $R_\odot$ and $1.3\,R_\odot$ it is $0^\circ.7$, while for $r>1.7\,R_\odot$ it is $2^\circ.8$. During steady-state convergence we apply an additional refinement with an angular resolution of $1^\circ.4$ at the heliospheric current sheet. The final total number of computational cells is 23,236,608. In addition to the high radial resolution near the Sun, we artificially broaden the transition region, similar to that described in \citet{Lionello:2009,vanderHolst:2014} to be able to resolve that region.

\section{Results}
\label{results}

The simulation was performed with protons, electrons, and charge states of all oxygen ions. However, in the following, we will analyze only the results of protons and the most abundant oxygen ions O$^{5+}$, O$^{6+}$ and O$^{7+}$. The abundance of oxygen is in the lower transition region (50,000 K) set to values provided by the CHIANTI database, resulting in an oxygen-to-proton abundance ratio of 0.0007762.

\begin{figure}
    \centering
    \includegraphics[width=0.49\linewidth]{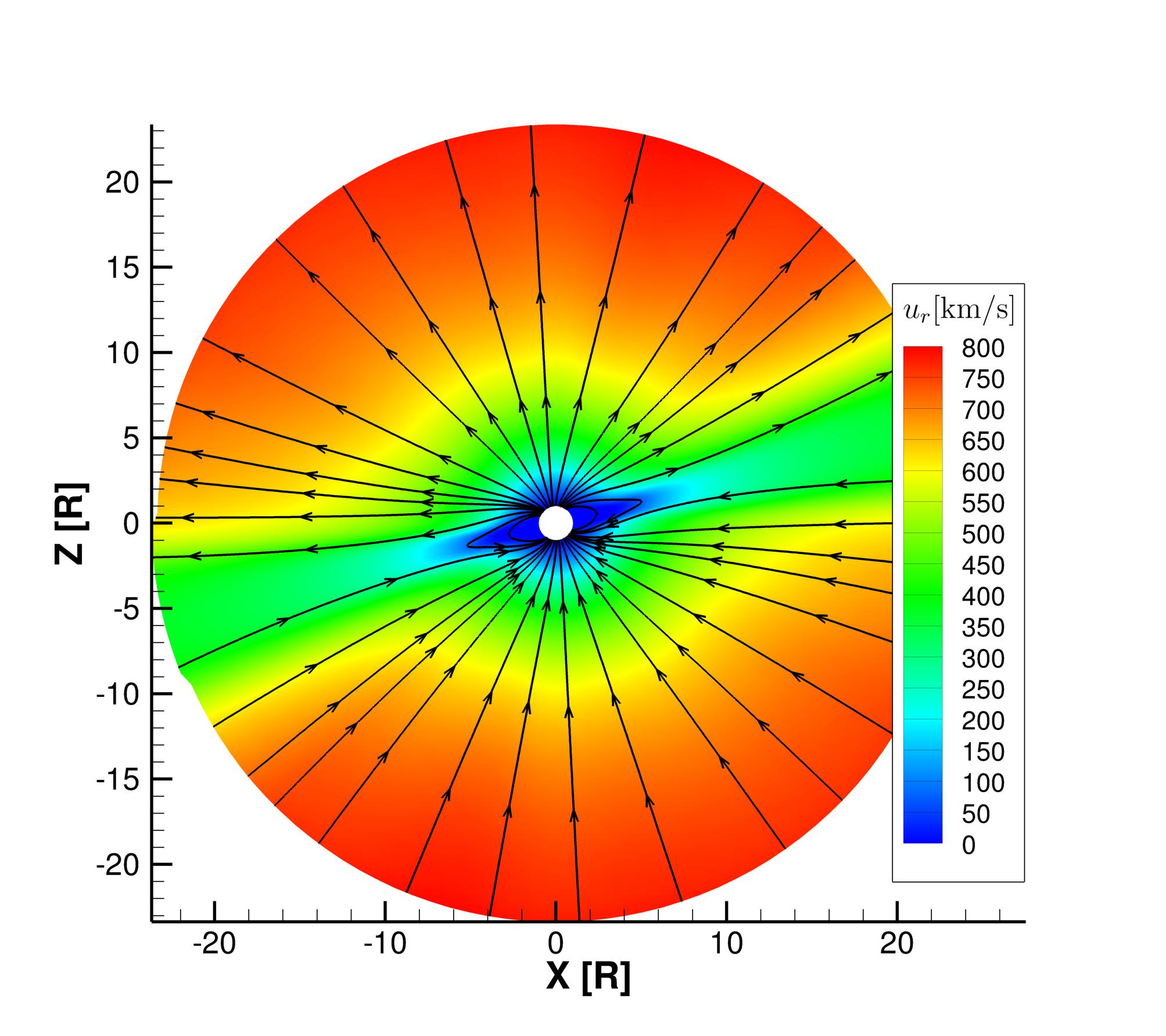}
    \includegraphics[width=0.49\linewidth]{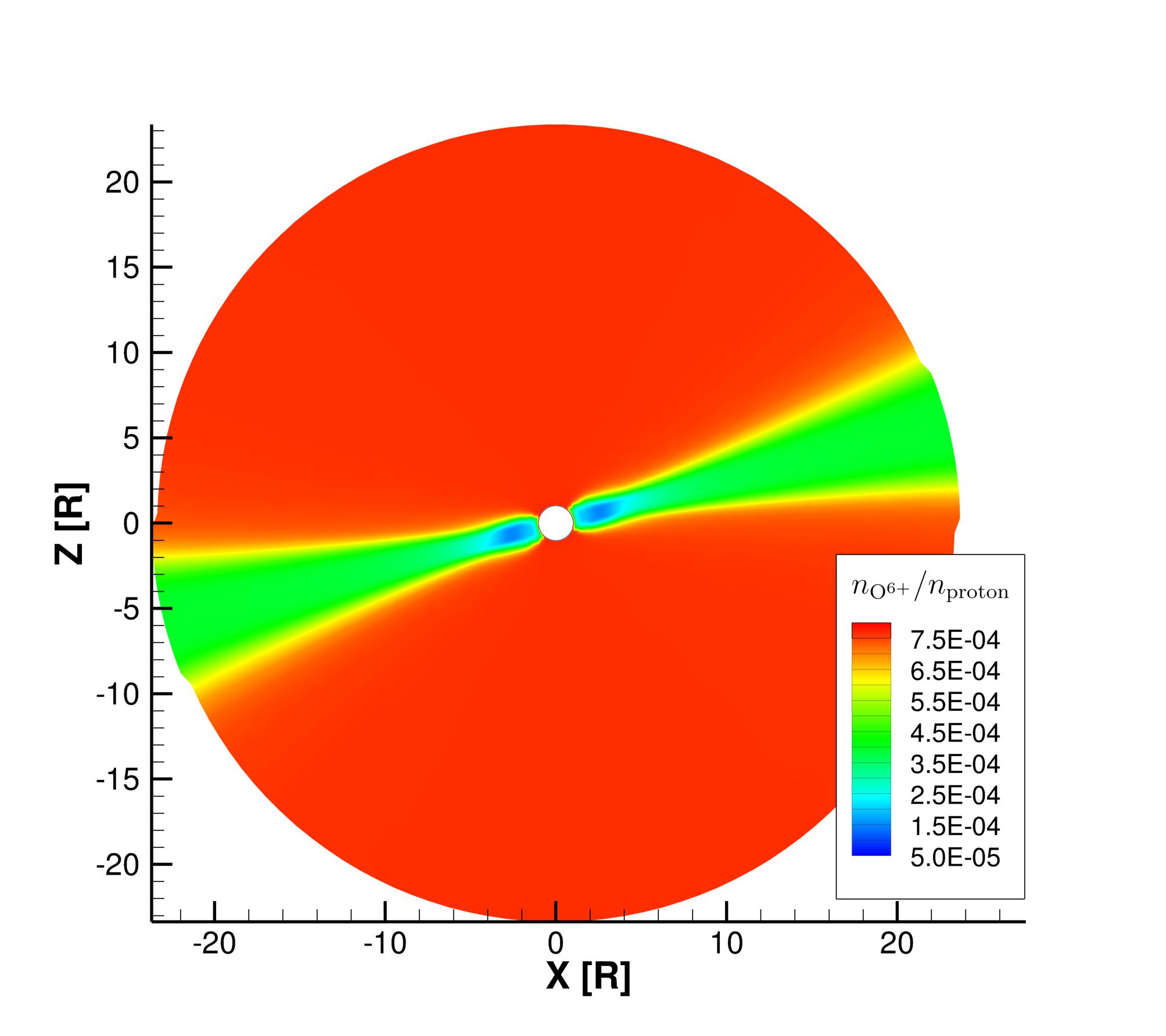}\\
    \vspace{-0.3cm}
    \includegraphics[width=0.49\linewidth]{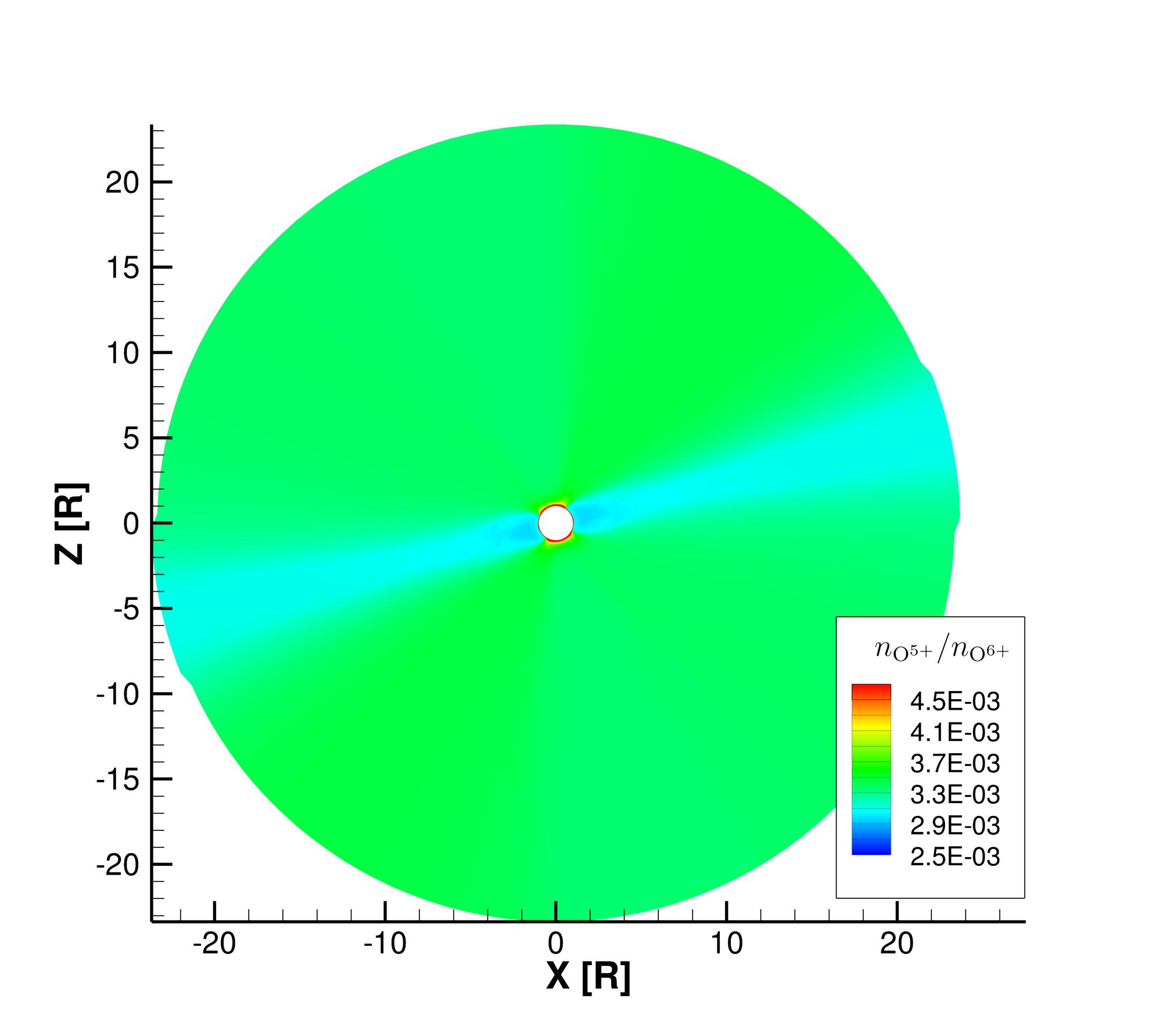}
    \includegraphics[width=0.49\linewidth]{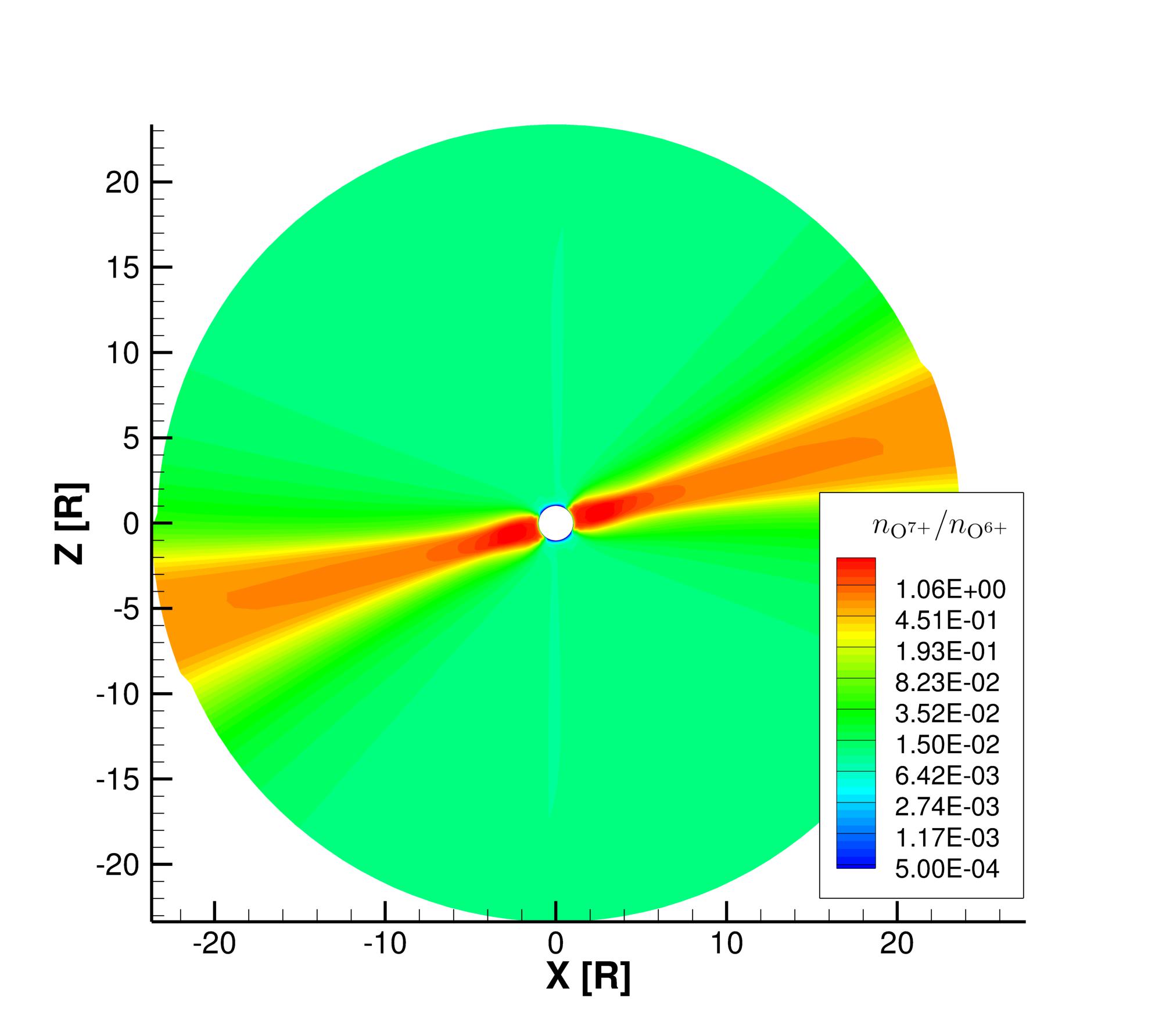}
    \caption{Meridional slice ($Y = 0$ plane) of the solar corona showing in color contour the radial velocity $u_r$ (top left), $n_{\rm O^{6+}}/n_{\rm proton}$  density ratio (top right), $n_{\rm O^{5+}}/n_{\rm O^{6+}}$ density ratio (bottom left), $n_{\rm O^{7+}}/n_{\rm O^{6+}}$ density ratio (bottom right). Streamlines represent field lines without the out-of-plane component.}
    \label{fig:density}
\end{figure}

We evolve the solar wind model through local time stepping to steady state. In Figure \ref{fig:density}, we show the solar wind solution in a meridional slice. The radial velocity in the color contour in the top-left panel shows the fast wind with a speed of almost 800\,km/s and the slow wind with a speed of 360\,km/s. A heliospheric current sheet is formed at the top of the helmet streamer that separates outward pointing field lines of opposite direction. The density ratio $n_{\rm O^{6+}}/n_{\rm proton}$ shows a similar bimodal structure as the wind speed with a ratio of $3.93\times10^{-4}$ in the slow wind and a ratio of $7.66\times10^{-4}$ in the fast wind near the outer boundary at 21\,R$_\odot$. The density ratio of $n_{\rm O^{5+}}/n_{\rm O^{6+}}$ is smaller, with a ratio of $3.4\times10^{-3}$ in the fast wind and $3.1\times10^{-3}$ in the slow wind, in line with measurements taken both by Ulysses/SWICS and ACE/SWICS in polar and equatorial coronal holes. The density ratio $n_{\rm O^{7+}}>n_{\rm O^{6+}}$ is also lower in the fast solar wind with a ratio of $\approx 0.01$ in the fast wind -- again in agreement with measured values -- and $0.69$ in the slow wind, more in line with the slow wind measured during solar maximum (\citet{Lepri:2013}). In the helmet streamer $n_{\rm O^{7+}}>n_{\rm O^{6+}}$, except for $r\lesssim1.15\,$R$_\odot$.

The bimodal structure in the charge state ratio is in line with the expectation that the $n_{\rm O^{7+}}/n_{\rm O^{6+}}$ ratio is higher in hotter, denser, and slower moving plasma, whereas the $n_{\rm O^{5+}}/n_{\rm O^{6+}}$ ratio is expected to be lower in those regions. See \citet{Szente:2022} for more details on the charge state ratios in the solar wind.

\begin{figure}
    \centering
    \includegraphics[width=0.49\linewidth]{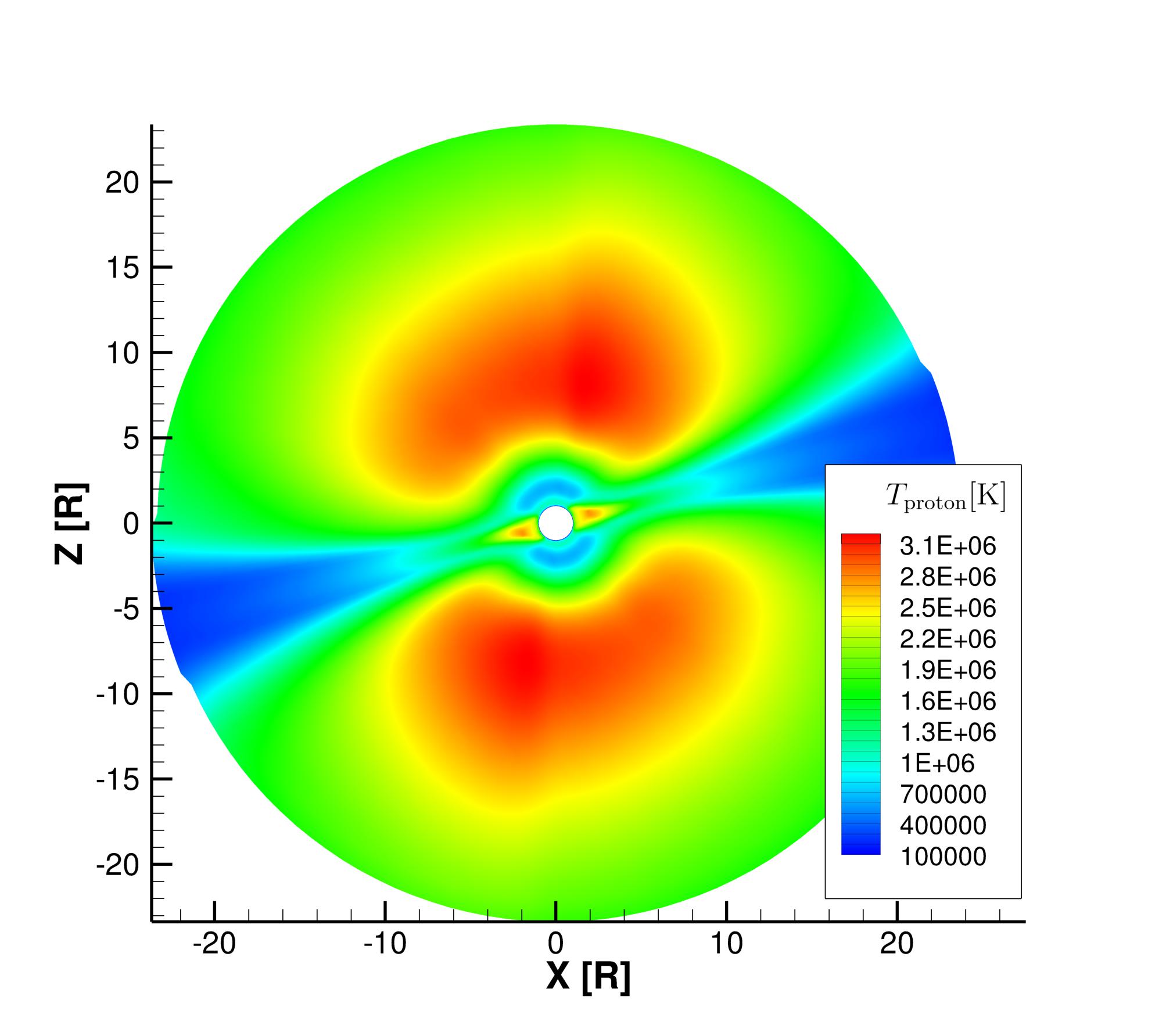}
    \includegraphics[width=0.49\linewidth]{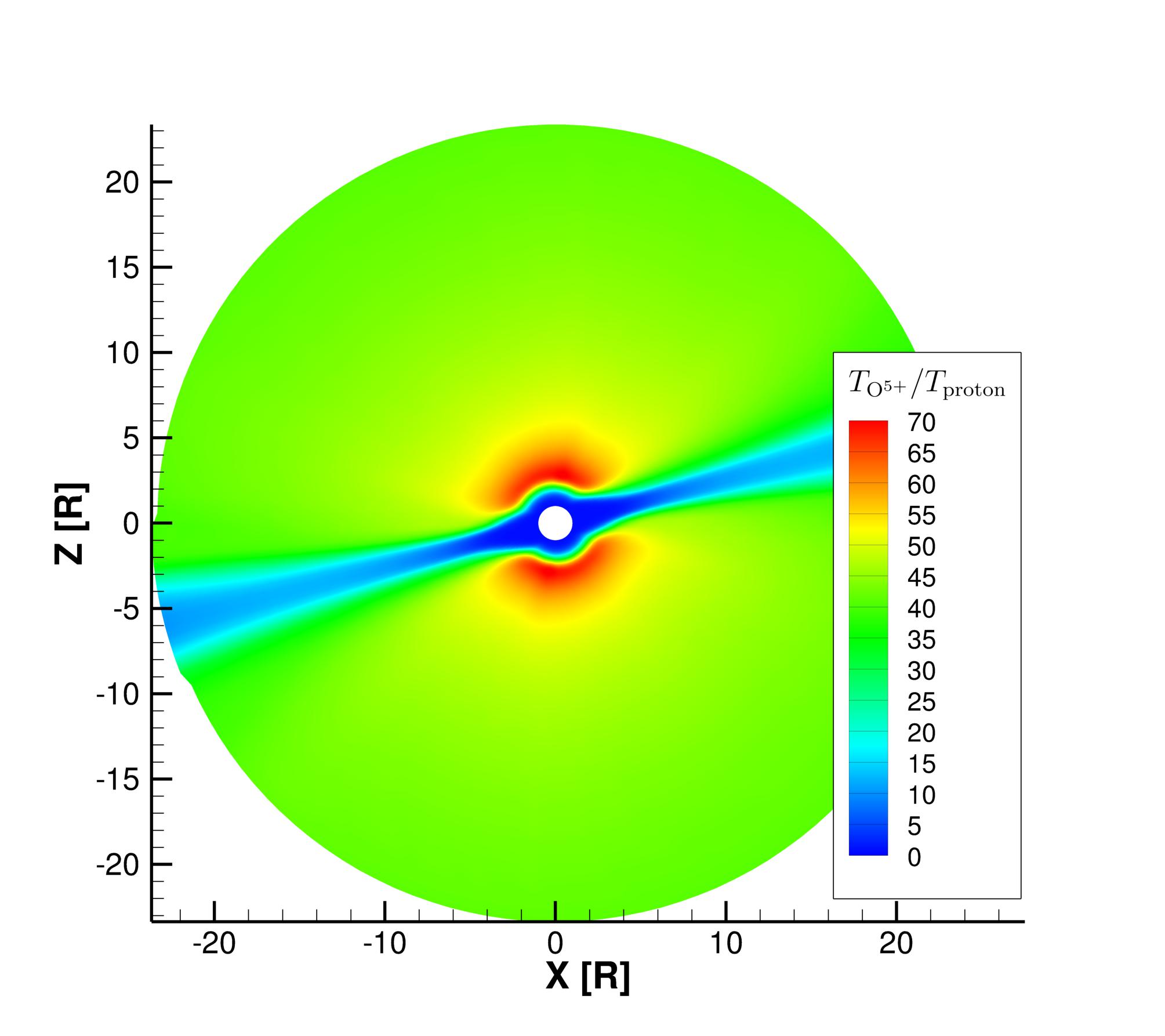}\\
    \vspace{-0.3cm}
    \includegraphics[width=0.49\linewidth]{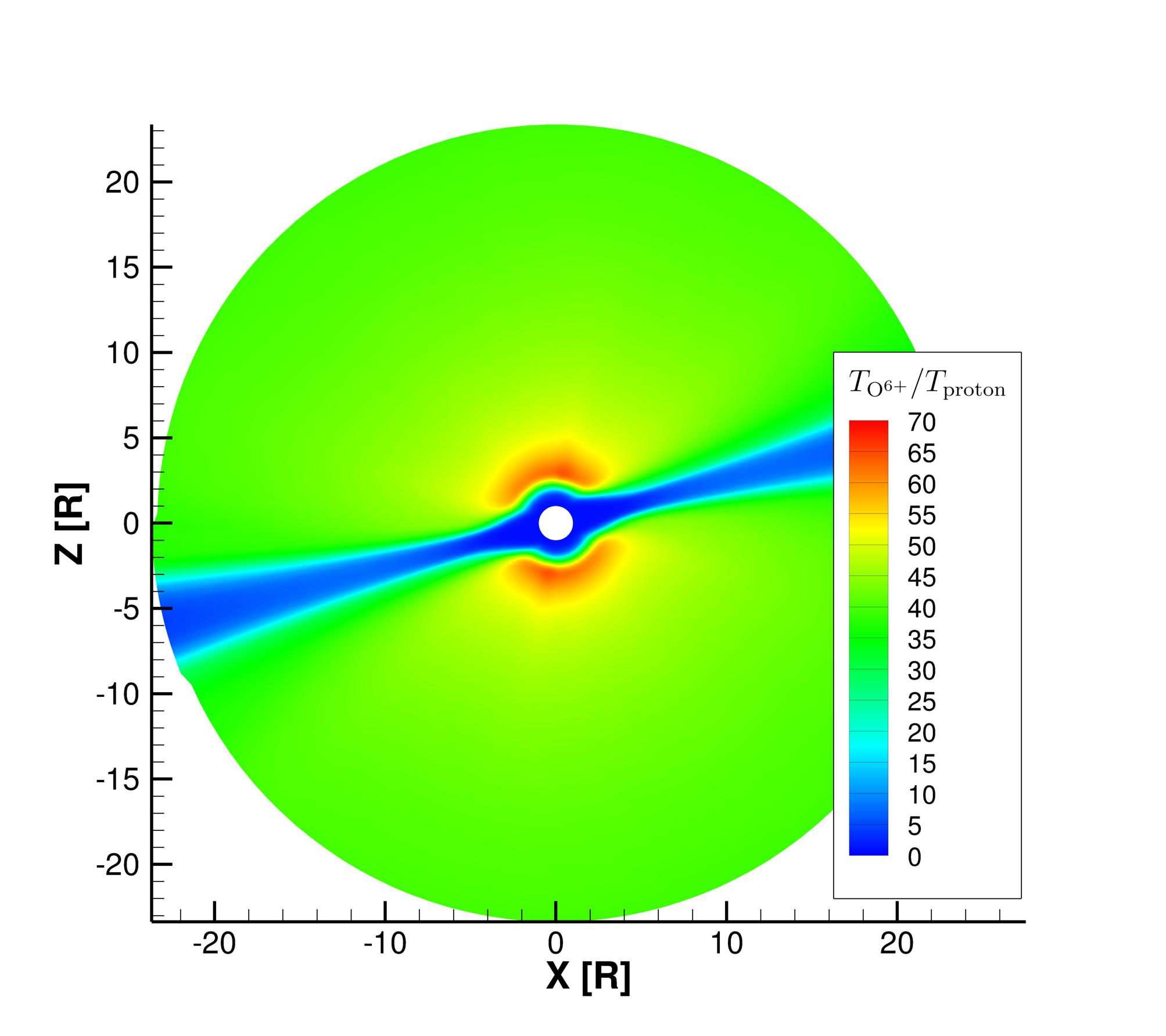}
    \includegraphics[width=0.49\linewidth]{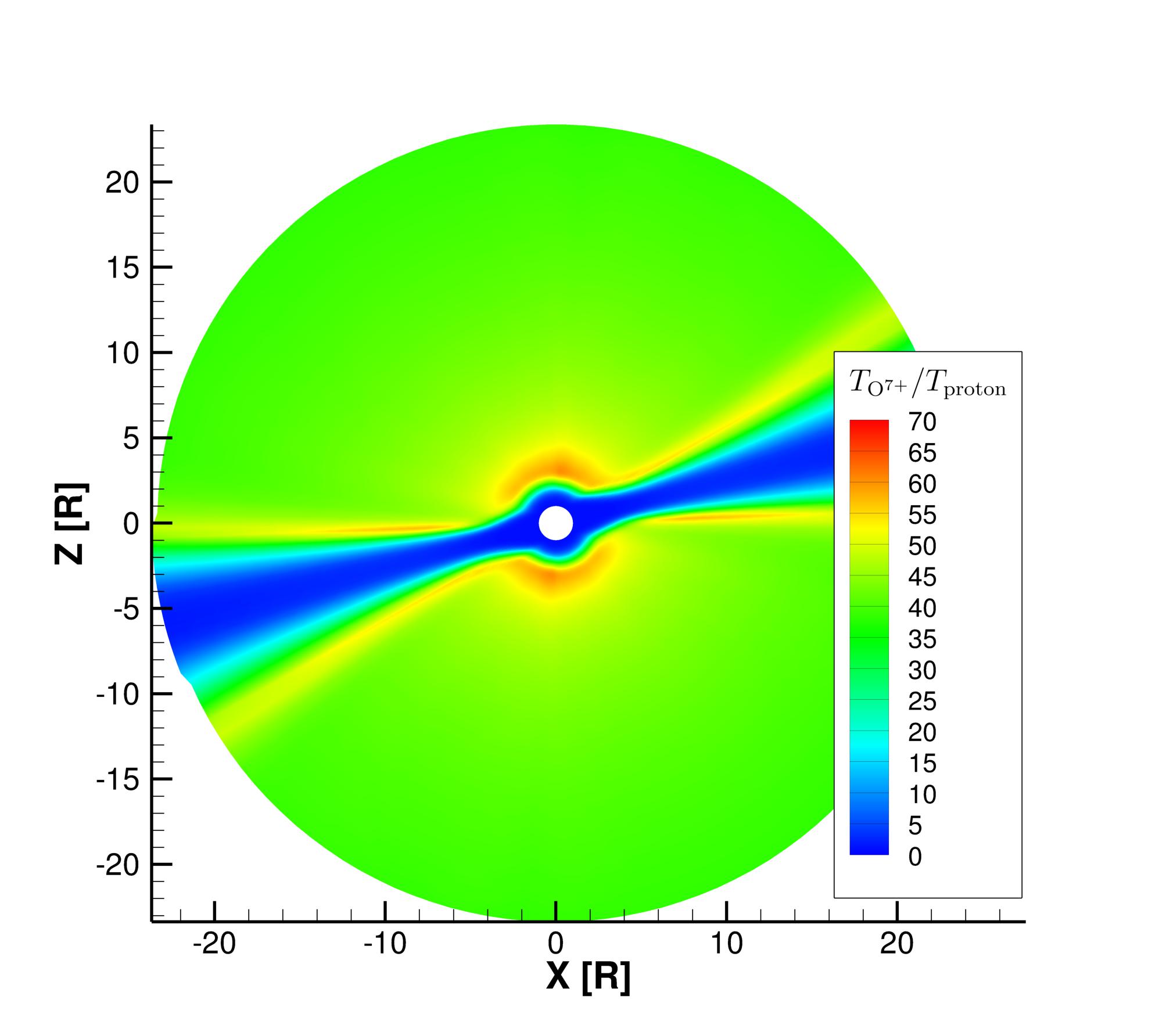}
    \caption{Meridional slice ($Y = 0$ plane) of the solar corona showing in color contour the proton temperature $T_{\rm proton}$ (top left), $T_{\rm O^{5+}}/T_{\rm proton}$  temperature ratio (top right), $T_{\rm O^{6+}}/T_{\rm proton}$ (bottom left), and $T_{\rm O^{7+}}/T_{\rm proton}$ (bottom right).}
    \label{fig:temperature}
\end{figure}

The ion temperatures are depicted in a meridional slice in Figure \ref{fig:temperature}. The temperatures display a bimodal structure, with highest temperatures in the fast solar wind and lower temperatures in the slow wind. For protons (top-left panel), the maximum temperature in the fast wind is about $3\times10^6\,$K and occur at around 10~R$_\odot$, and then they decrease beyond that distance. On the contrary, the highest proton temperature in the helmet streamer is $T_{\rm proton}\approx 2.75\times10^6\,$K and occurs in the innermost corona within 1~R$_\odot$ from the photosphere, only to decrease at larger distances. The O$^{5+}$ to proton temperature ratio in the top-right panel shows that except for the regions closest to the photosphere the temperatures of all Oxygen ions are much larger than the proton temperatures, and that differences are highly dependent on wind type and distance. The largest differences occurr within 5~R$_\odot$ and reach a maximum of 67.9 at $r=2.95\,$R$_\odot$ in the polar direction. At $r=21\,$R$_\odot$ this ratio reduces to 42 in the fast wind, while it is 11.5 in the slow wind. For the O$^{6+}$ to proton temperature ratio in the bottom-left panel, these values are a little bit smaller, the maximum is 61.8 at $r=3.1\,$R$_\odot$ in the polar direction. At $r=21\,$R$_\odot$ this ratio is 40 in the fast wind, while it is 5.2 in the slow wind. For O$^{7+}$ in the bottom-right panel, this is even smaller with a maximum value of 58 at $r=3.15\,$R$_\odot$ in the polar direction. At $r=21\,$R$_\odot$ this ratio is 39 in the fast wind, while it is 1.96 in the slow wind. Throughout the inner heliosphere, it thus looks like Oxygen is heated more than mass-proportionally almost everywhere, except in the narrow region of the slow wind where the heating is less than mass proportional. The ratios are all 1.0 in the helmet streamer.

\begin{figure}
    \centering
    \includegraphics[width=0.5\linewidth]{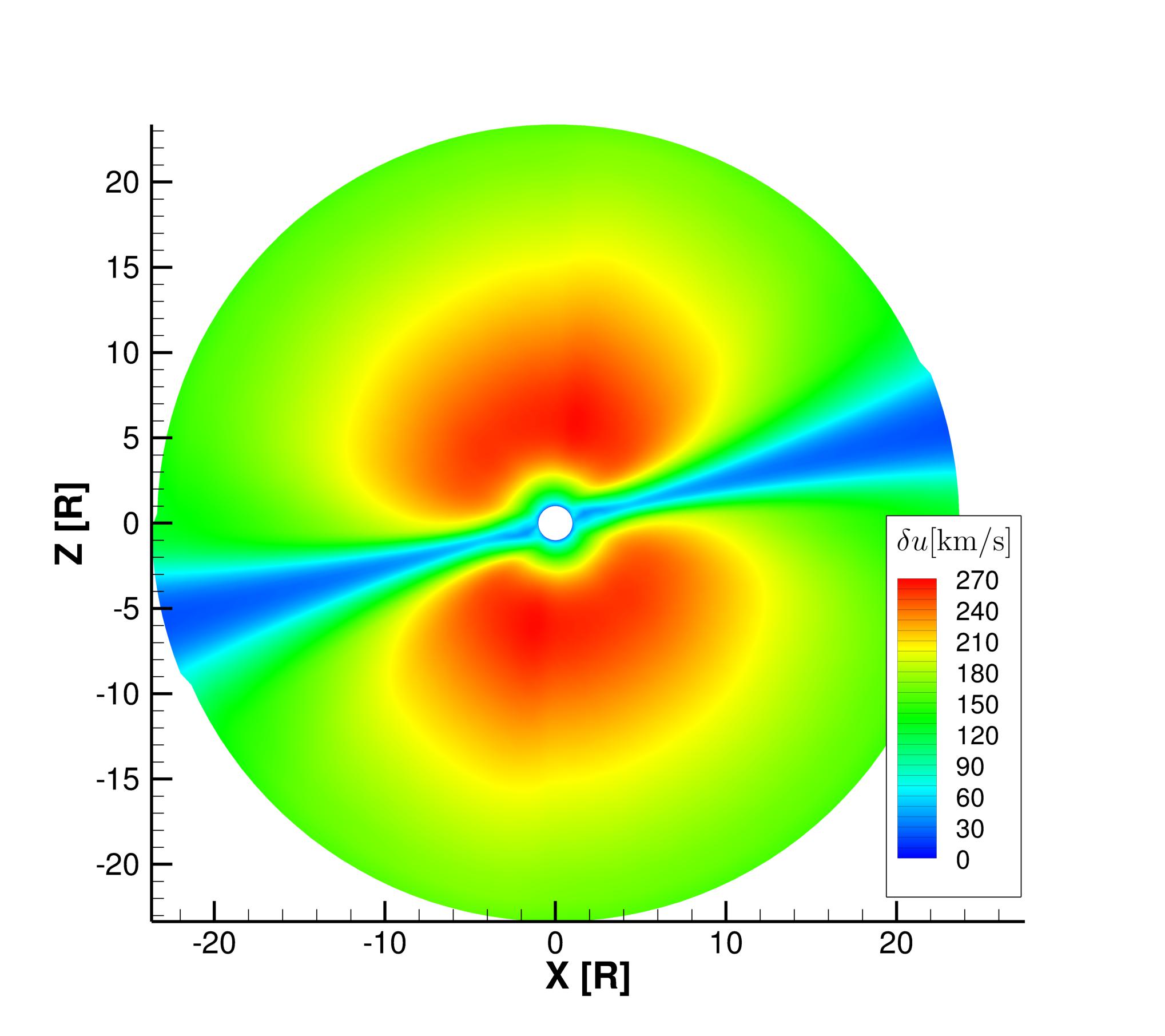}
    \caption{Meridional slice ($Y = 0$ plane) of the solar corona showing in color contour the amplitude of the turbulent velocity fluctuations $\delta u = \sqrt{(w+w_D)/\rho}$.}
    \label{fig:deltau}
\end{figure}

Figure \ref{fig:deltau} shows in a meridional slice the amplitude of turbulent velocity fluctuations. The perpendicular ion heating, due to stochastic heating processes, dominates where the velocity fluctuations are significant. This is in the corona away from the Sun and the heliospheric current sheet (HCS) and results in significant ion temperatures in the fast wind and lower ion temperatures in the slow wind. Very close to the Sun and near the HCS, the electron heating is important, while the parallel proton heating is only significant very close to the HCS where the plasma beta $\beta_p=2\mu_0 p_p/B^2$ is high. We note that we did not include the parallel heating for the heavy ions, which might have an impact very close to the HCS. 

\subsection{Comparison with observations}
\label{observations}

\begin{figure}
    \centering
    \includegraphics[width=0.49\linewidth]{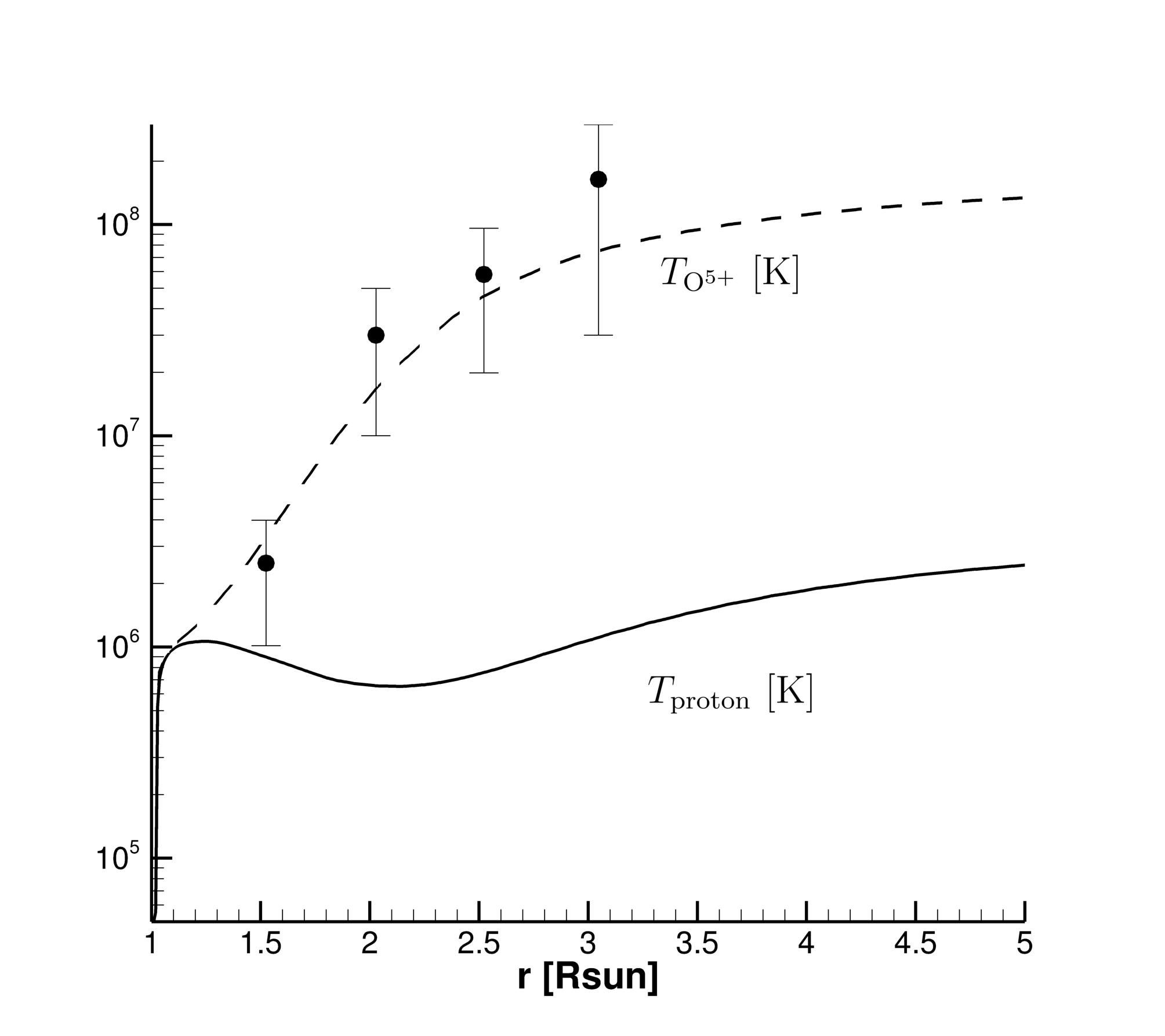}
    \caption{O$^{5+}$ (dashed line) and proton (drawn line) temperatures along the magnetic north axis. Observations from UVCS of O$^{5+}$ temperatures in the polar coronal hole from \citet{Rivera:2025} are included.}
    \label{fig:UVCS}
\end{figure}

In Figure~\ref{fig:UVCS}, we compare the temperature of O$^{5+}$ along the magnetic north axis with those obtained from UltraViolet Coronagraph Spectrometer \citep[UVCS,][]{Kohl:1995} on the Solar and Heliospheric Observatory (SOHO) satellite. These measurements were taken from \citet{Rivera:2025} (Figure 3) and result from the empirical model developed by \citet{Cranmer:2008}. The latter authors utilized several observations of polar coronal holes taken during the minimum of solar cycle 23 (1996-1997) to build empirical models of the plasma conditions that allow the calculation of the O$^{5+}$ line intensity and profile that best fit the observations. This procedure consists in exploring the space of the parameters involved in O$^{5+}$ line profile formation, and is made necessary by the non-trivial dependence of the O$^{5+}$ line formation on plasma parameters, which make a direct interpretation of line widths as ion temperatures impossible. However, this procedure dispenses from assumptions on coronal heating and acceleration, and utilizes only well-known concepts of radiative transfer in the context of line intensity formation so that the final results are relatively robust.  

Figure~\ref{fig:UVCS} shows that our model is capable of reproducing the very rapid rise of the O$^{5+}$ perpendicular temperature observed by UVCS, and that $r>2\,$R$_\odot$ the model results agree with the UVCS measurements well within the error bars. This is even more remarkable if we consider that the UVCS data correspond to observations between 1996 and 1997, while our simulation is for an idealized dipole solar wind case. 

For comparison with 1 AU data, the solar wind ion temperatures that we used come from the SWICS instrument \citep{Gloeckler:1998SSRv...86..497G} on board the ACE satellite. SWICS provided high quality measurements of heavy ion properties from the start of the mission in 1998 to 2011, after which an anomaly in the hardware increased the background and generated several invalid measurements. SWICS measured the bulk velocity,  thermal speed, and density for the most abundant ions in the solar wind: He$^{2+}$, C$^{4-6+}$, N$^{5-7+}$, O$^{5-7+}$, Ne$^{6-9+}$ (with the exception of Ne$^{7+}$, unavailable), and Fe$^{7-12+}$; we have used 2-hr averaged data. The thermal speed in this dataset only includes the component along the SWICS look direction.

The 1998--2011 time interval covers one full solar cycle, from the rising phase of cycle~23 to the rising phase of cycle~24. During this period, the magnetic configuration of the solar atmosphere underwent the typical massive changes from minimum to maximum; since we utilized a simplified photospheric magnetic model for our simulations (a simple dipole magnetic field), we needed to select data from a time period in the solar cycle where the solar magnetic field was closest to a dipole. The best choice is Carrington Rotation (CR) 2082 (5-Apr-2009 to 3-May-2009), during the deep minimum of the solar cycle~23.

As an example to test whether this model provided a reasonable estimate of ion temperatures, we focused on one of the most abundant elements in the solar wind: Oxygen, whose most abundant ions are O$^{5+}$, O$^{6+}$ and O$^{7+}$: the abundance ratios of these ions are routinely used to measure the freeze-in temperature of the solar wind (\citet{Landi:2012ApJ...758L..21L, Landi:2012ApJ...761...48L} and references therein) and the solar wind type (e.g. \citet{Zhao:2009GeoRL..3614104Z}). A more systematic comparison of predicted and measured solar wind ion temperatures will be the focus on the next paper of the series, utilizing a measured photospheric magnetic field map.

\begin{figure}
    \centering
    \includegraphics[width=0.49\linewidth]{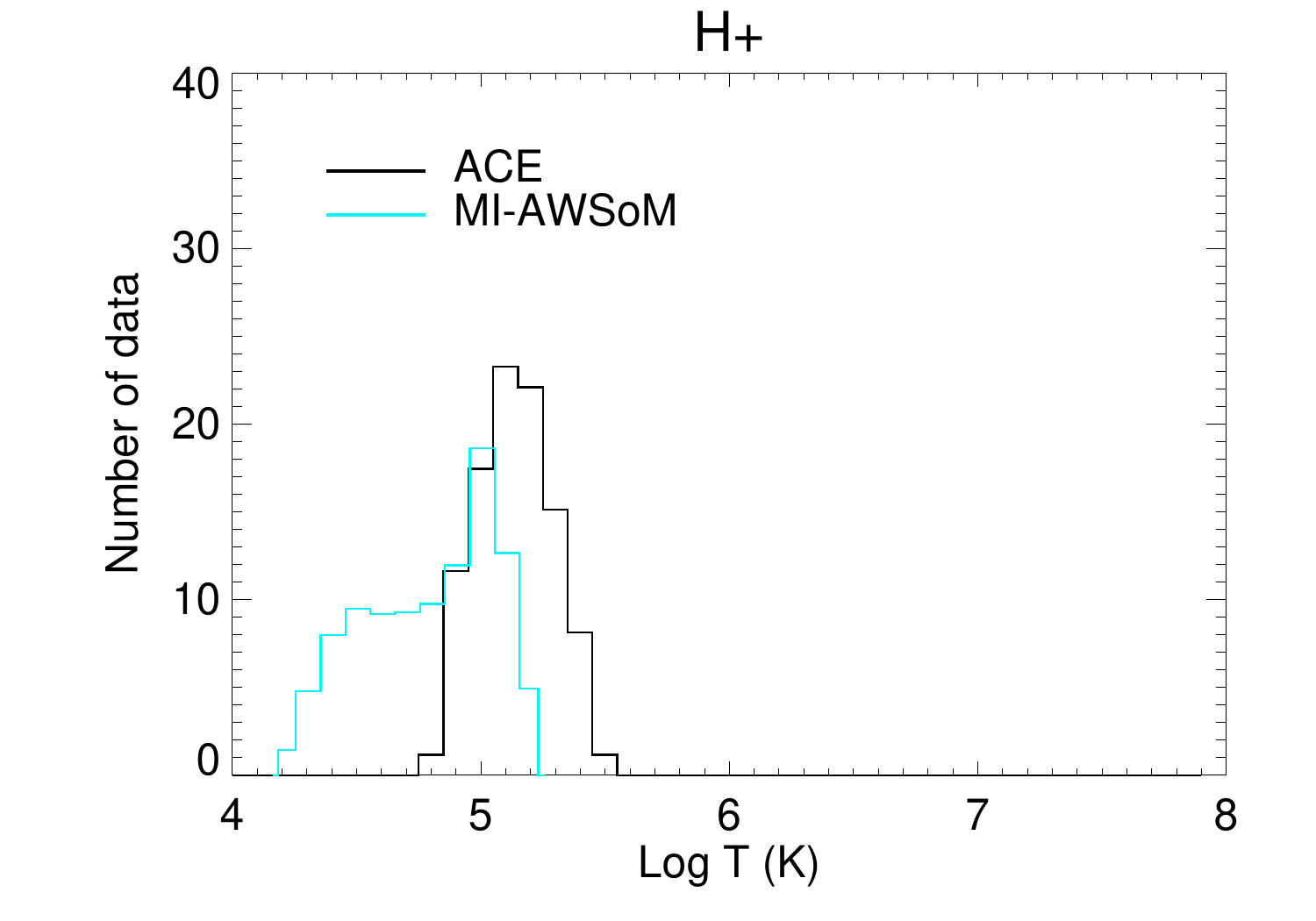}
    \includegraphics[width=0.49\linewidth]{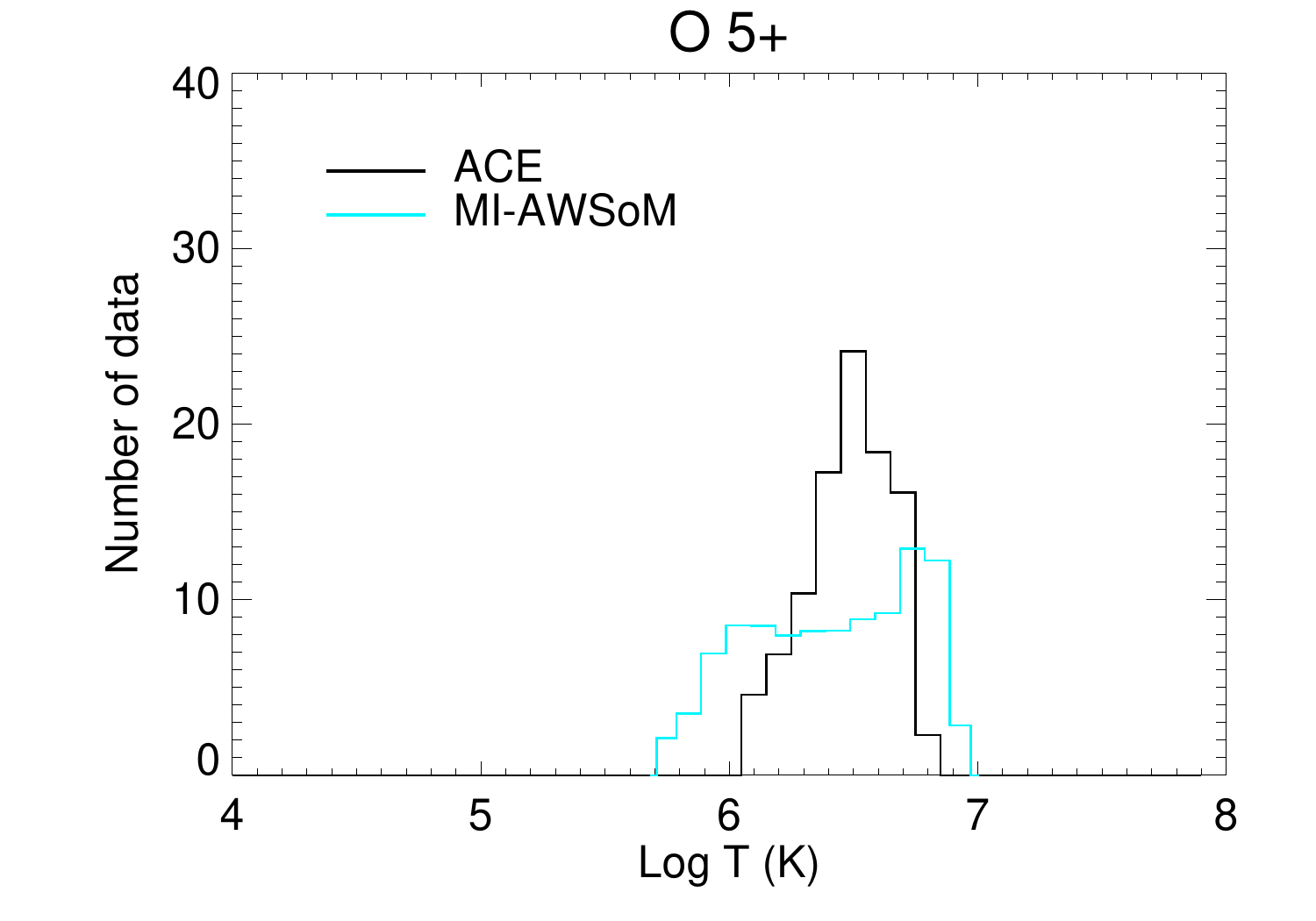}\\
    \includegraphics[width=0.49\linewidth]{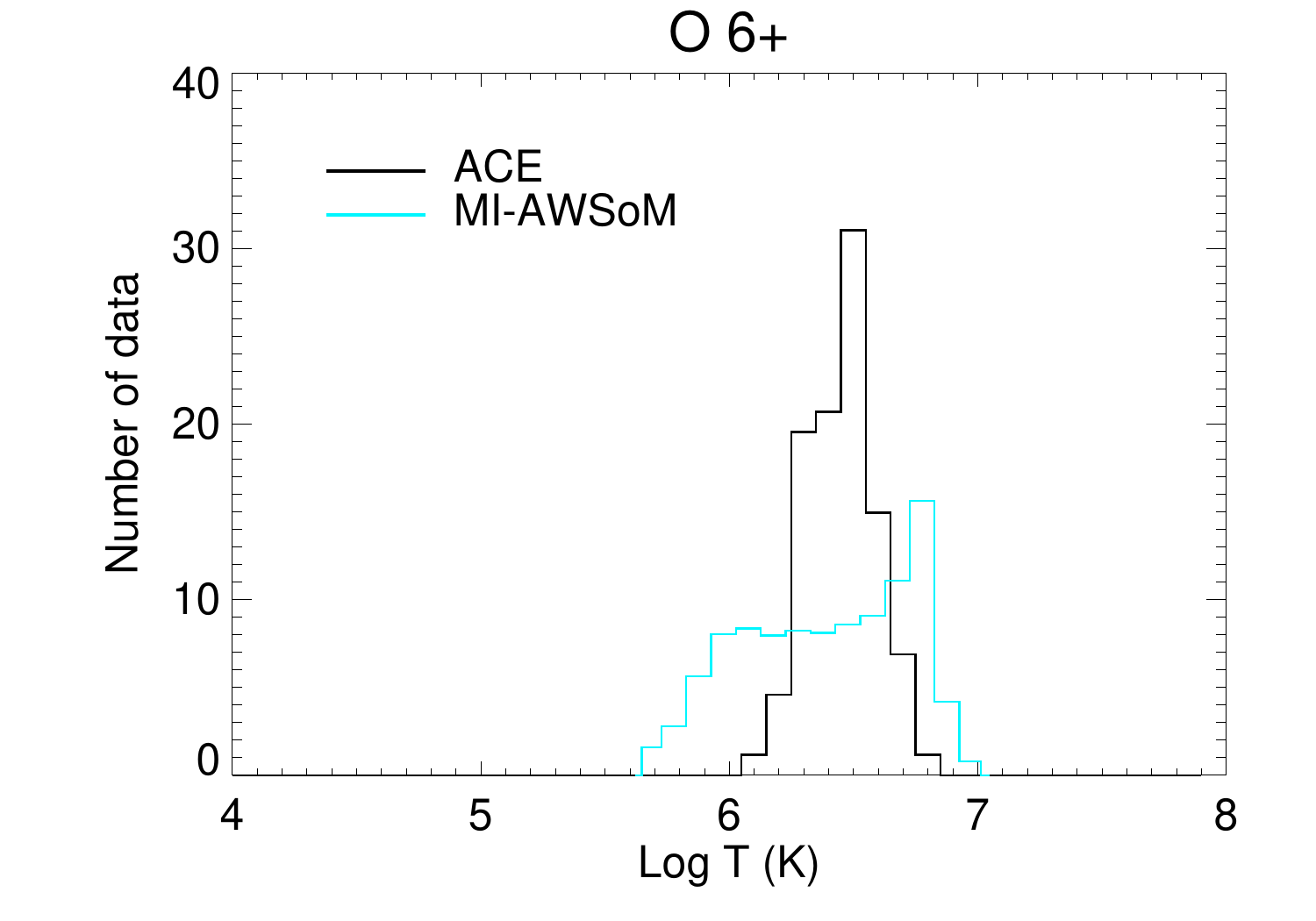}
    \includegraphics[width=0.49\linewidth]{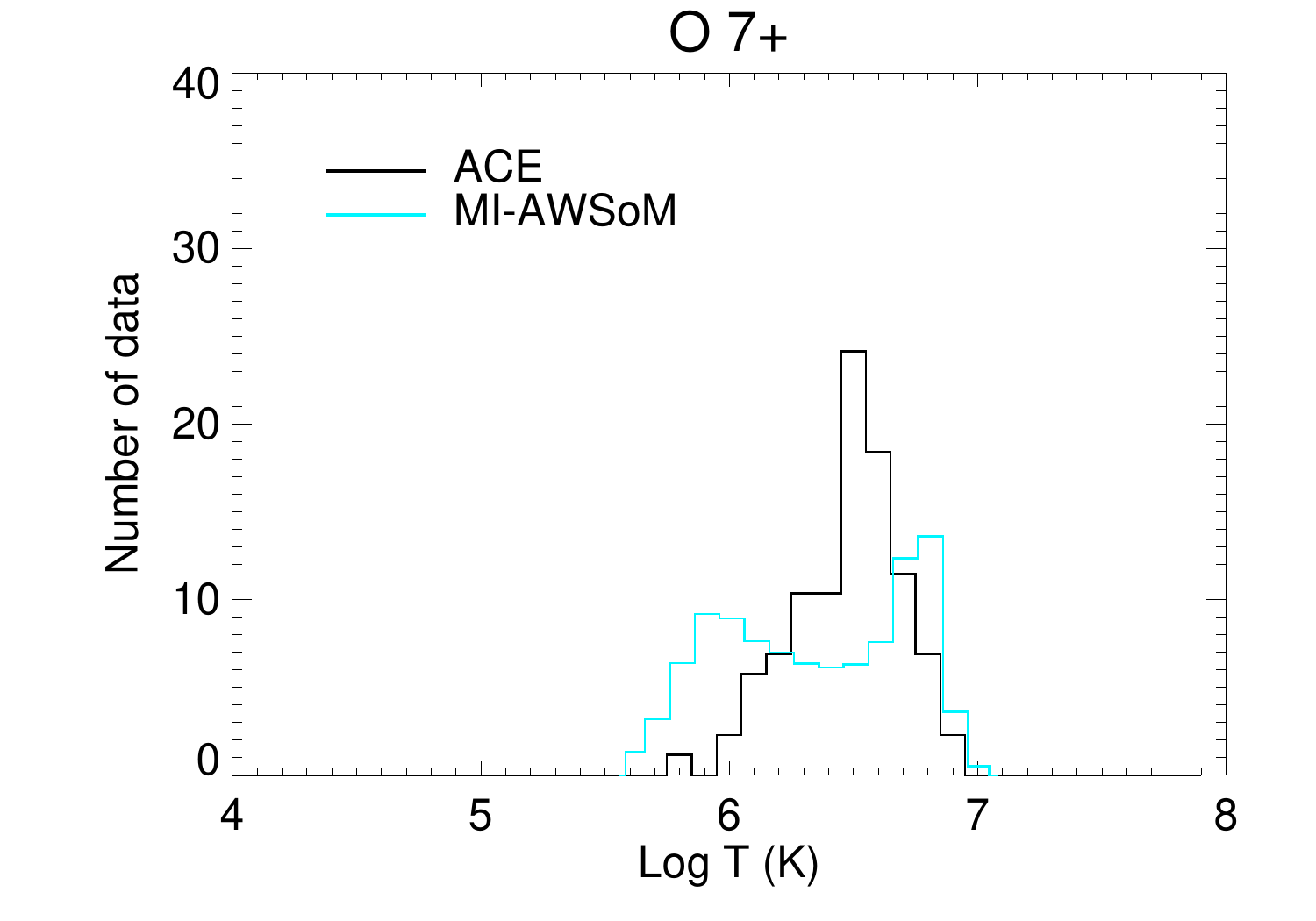}
    \caption{Comparison of the counts per temperature bin in the fast solar wind (beyond 500 km/s speed) at 1~AU [$\Delta \log T (\text{K}) = 0.1$] of the SWICS data (CR~2080 to CR~2084) to the model results for protons, O$^{5+}$, O$^{6+}$, and O$^{7+}$. The number of counts is normalized to 100 for both data and model to facilitate comparison.}
    \label{fig:ih}
\end{figure}

To improve the quality of the sample, for each ion, we utilized only time stamps with a number of counts larger than 15. Since we are only interested in the background solar wind, we have removed all data taken during ICME events, identified using the Richardson \& Cane list \citep{Cane:2003JGRA..108.1156C, Richardson:2010SoPh..264..189R}; we further excluded wind streams not present in that list, which were either faster than 800~km/s, or with Fe$^{16+}$/Fe ratio values larger than 0.1 or O$^{7+}$/O$^{6+}$ ratio values larger than 1. It turned out that during CR~2082 measurements of wind faster than 500~km/s were very scarce, so the number of suitable time stamps was extremely limited. In order to increase the number of suitable time stamps we used all suitable measurements with speed larger than 500~km/s measured from CR~2080 to CR~2084, for a total of five full CRs, spanning from 10-Feb-2009 to 27-Jun-2009, see Figure~\ref{fig:ih}. This down-selection of the intervals for the statistical analysis resulted in using $5.4\%$ of the dataset. We note that during the 2008-2009 minimum there was not much fast wind (that is, with speed larger than 500 km/s).

The predicted ion temperatures span a broader range of temperatures than the observed ones but are found at the same temperatures as the latter. The predicted values are clustered around a peak at around 5-6 MK and a second smaller peak at lower temperatures ($\approx$1MK); on the contrary, the observed values are clustered around a narrower peak of $\approx$ 3MK. This comparison clearly shows that the model, though the simplification of a simplistic dipole magnetic field configuration, is nonetheless capable of reproducing the observed values.

\section{Summary}
\label{summary}

We have presented a three-dimensional multi-ion solar wind model. This model includes separate mass densities and temperatures for each ion, but differential streaming is excluded for now. To address coronal heating and solar wind acceleration, we use low-frequency, reflection-driven incompressible turbulence. The energy partitioning of the turbulence dissipation is based on stochastic heating and linear Landau- and transit time damping. Currently, only the stochastic heating mechanism is heating the minor ions. We have included ionization-recombination processes to obtain charge states, similar to \citet{Szente:2022}. A more rigorous account of the effects of ionization-recombination on the ion pressure equations is diverted to future work.

We performed a tilted magnetic field dipole test with protons, electrons, and oxygen charge state. The simulation shows that O$^{6+}$ is the dominant oxygen ion, while in the slow wind O$^{7+}$ number density is almost as high as that of O$^{6+}$. In the fast wind, the oxygen ions are heated more than mass proportional, while in the slow solar wind the heating is less than mass proportional. The higher oxygen temperature in the fast wind is a direct result of the higher turbulence level resulting in our model in increased stochastic heating in those regions. For the oxygen charge state ratio, our model shows the expected increase in $n_{\rm O^{7+}}/n_{\rm O^{6+}}$ in higher in hotter, denser, and slower moving plasma, while a lower $n_{\rm O^{5+}}/n_{\rm O^{6+}}$ in those regions. We compared the idealized simulation results with SOHO/UVCS. The simulated O$^{5+}$ temperatures are well within the error bars of the observations. Additionally, we compared the oxygen temperature with ACE/SWICS and find that the model is qualitatively in agreement with the data for the fast wind. For the slow wind comparison, we will need to model with magnetic maps for the various Carrington rotations.

In future simulations, more realistic background can be obtained by using magnetic field maps. We can then compare those results with the spectral data using the SPECTRUM model of \citet{Szente:2019,Szente:2023}. The multi-ion solar wind model also needs further improvements by adding differential streaming as well as linear Landau and transit time damping for the minor ions using the PLUME model \citet{Klein:2015}. These model improvements will be reported in a follow up paper.

\section{Acknowledgments}
EL and JSz were supported by NASA Awards 80NSSC22K0750, 80NSSC21K1124. \\
SOHO is a project of international cooperation between ESA and NASA.\\
We thank the ACE SWICS instrument team and the ACE Science Center for providing the ACE data.\\
We acknowledge the developers of the Space Weather Modeling Framework and the Center for Space Environment Modeling at the University of Michigan. This work was carried out using and further developing the Space Weather Modeling Framework.\\
Resources supporting this work were provided by the NASA Advanced Supercomputing (NAS) Division at Ames Research Center.

\bibliography{references}{}
\bibliographystyle{aasjournalv7}

\end{document}